\documentclass[12pt]{article}
\usepackage{amsmath,amsfonts,amssymb,dcolumn,lscape,graphicx,subfigure}
\usepackage{color}
\begin{document}
\baselineskip=5.5mm
\newcommand{\be} {\begin{equation}}
\newcommand{\ee} {\end{equation}}
\newcommand{\Be} {\begin{eqnarray}}
\newcommand{\Ee} {\end{eqnarray}}
\newcommand{\tmun}[1]{\textcolor{green}{#1}}
\newcommand{\tros}[1]{\textcolor{blue}{#1}}
\renewcommand{\thefootnote}{\fnsymbol{footnote}}
\def\a{\alpha}
\def\b{\beta}
\def\g{\gamma}
\def\G{\Gamma}
\def\d{\delta}
\def\D{\Delta}
\def\e{\epsilon}
\def\k{\kappa}
\def\l{\lambda}
\def\L{\Lambda}
\def\t{\tau}
\def\om{\omega}
\def\Om{\Omega}
\def\s{\sigma}
\def\lg{\langle}
\def\rg{\rangle}
\begin{center}
{\Large {\bf Aging effects manifested in the potential energy landscape of a model glass former} }\\
\vspace{0.5cm}
\noindent
{\bf Christian Rehwald$^1$, Nicoletta Gnan$^2$, Andreas Heuer$^1$,
Thomas Schr\o der$^2$, Jeppe C. Dyre$^2$ and 
Gregor Diezemann$^3$} \\
{\it
$^1$Institut f\"ur Physikalische Chemie, Universit\"at M\"unster,
Corrensstr. 30, 48149 M\"unster, FRG\\
$^2$Department of Sciences, DNRF Centre 'Glass and Time', IMFUFA, Roskilde University, 
P.O. Box 260, DK-4000 Roskilde, Denmark\\ 
$^3$Institut f\"ur Physikalische Chemie, Universit\"at Mainz,
Welderweg 11, 55099 Mainz, FRG
\\}
\end{center}
\vspace{1cm}
\noindent
{\it
We present molecular dynamics simulations of a model glass-forming liquid (the binary Kob-Anderson Lennard-Jones model) and consider the distributions of inherent energies and metabasins during aging.
In addition to the typical protocol of performing a temperature jump from a high temperature to a low destination temperature, we consider the temporal evolution of the distributions after an 'up-jump', i.e. from a low to a high temperature.
In this case the distribution of megabasin energies exhibits a transient two-peak structure.
Our results can qualitatively be rationalized in terms of a trap model with a Gaussian distribution of trap energies.
The analysis is performed for different system sizes.
A detailed comparison with the trap model is possible only for a small system because of major averging effects for larger systems.
\\
}
\section*{I. Introduction}
The primary relaxation of supercooled liquids still poses challenging questions regarding its mechanism, and a detailed theoretical understanding of the often dramatic temperature dependence of the relaxation time is still lacking. 
In the past decade it has been shown that the $\a$-relaxation has to be viewed as dynamically heterogeneuous\cite{Sillescu1999, Ediger2000}.

The present paper focuses on one particular aspect of the primary relaxation, namely the response of the system to an abrupt change in temperature.
The out-of-equilibrium dynamics of glassy systems has been studied for a long time, 
for a review see\cite{Bouchaud:1998}. 
Many investigations were devoted to the study of violations of the fluctuation-dissipation theorem\cite{Crisanti:2003p3550}. 
In particular, the question regarding a meaningful definition of an effective temperature remains to be answered as there are still conflicting results (for a recent investigation see\cite{JabbariFarouji:2008p4149} and the literature therein).
Apart from the theoretically challenging possibility of a description of the off-equilibrium dynamics in terms of an effective thermodynamics, the aging behavior of glassy systems is of huge interest, both theoretically and for practical reasons. 
Accordingly, it has been studied for a long time.
In particular, the rotational and translational dynamics has been investigated with the result that the aging properties of glass-forming liquids are determined by the primary relaxation, cf. 
ref.\cite{Thurau:2002p4151, Lunkenheimer:2005p4153}, albeit there are still some discussions about the detailed analysis of the corresponding data.

In addition to these investigations of the aging behavior, also the interesting question regarding the detailed interrelation of the aging properties of supercooled liquids and the nature of the dynamic heterogeneities is yet to be answered.
In the simplest conceivable picture the heterogeneous dynamics, as well as the aging properties, are governed by a distribution of activation energies.
Various theories and models of the glass transition deal with such distributions in one way or the other.
A detailed analysis of the temperature- and time-dependent properties of this distribution should be helpful in discriminating among different models.

If computer simulations on models for viscous liquids are considered, one has to face the problem of identifying a relevant distribution of activation energies.
Even though this point does not appear to be resolved completely, there seems to be growing evidence that the inherent structure (IS) energies or - in a more detailed picture - the metabasin (MB) energies are relevant for the determination of the activation energies\cite{Heuer:PETReview}.
This view partly emerges from the comparison of the results of molecular dynamics simulations with the predictions of a simple trap model with a Gaussian density of 
states\cite{DYRE:1995p2812}.
This Gaussian trap model qualitatively captures some relevant features of the energy landscape of simulated glassy systems\cite{Heuer:PETReview,DYRE:1995p2812,Denny:2003p3738}.
The comparison can be made more quantitative if a slightly extended version of the trap model is employed in the analysis\cite{HDS2005}.
From this analysis it appears that the IS- or MB-energies are intimately related to the relevant activation energies.

The aging behavior of the distribution of IS-energies of the Kob-Andersen binary mixture Lennard Jones (KABLJ) model\cite{KA1995a,KA1995b} was investigated by Saika-Vovoid and Sciortino\cite{Sciortino2004}.
These authors found that after a quench from a high temperature to some low temperature the mean SI-energy drops continuously and monotonously from its initial to the final value.
Furthermore, the width of the distribution of IS-energies decreases and then increases again as a function of the time that has elapsed after the quench.
An analysis of the aging behavior of the distribution of the energies in the Gaussian trap model shows a very similar behavior, cf. ref.\cite{DYRE:1995p2812,G56}.
Also the single-particle dynamics of a simulated glass-forming liquid immediately after a temperature   jump are fully compatible with the predictions of the Gaussian trap model\cite{Rottler2009}.

For a more quantitative analysis of the dynamics of simulated supercooled liquids the use of MBs rather than ISs has proven advantageous.
The reason is that IS trajectories are dominated by a large and temperature-dependent fraction of correlated forward-backward jumps.
In contrast, the MB trajectories can be interpreted in terms of a random-walk in configuration space with a temperature independent jump-length\cite{Heuer:PETReview,Doliwa2003}.
The complexity of the dynamics, giving rise, e.g., to the existence of dynamic heterogeneities, is fully reflected by the broad waiting time distribution.
This picture of a broad distribution of independent random jumps among different configurations presents the physical basis of the trap model.

In the present paper we study the temporal evolution of the IS-energies and the MB-energies of the KABLJ\cite{KA1995a,KA1995b} system after temperature jumps.
For a jump from a high to a low temperature, our results closely resemble those obtained by 
Saika-Vovoid and Sciortino\cite{Sciortino2004}.
Motivated by the finding that for the trap model after an 'up-jump' in temperature the energy-distribution exhibits a two-peak structure\cite{DYRE:1995p2812}, we show that this feature is also found in MD-simulations.
We discuss our findings with particular emphasis on their relevance to possible experimental realizations.

After a brief review of the aging-properties of the energy distributions in the Gaussian trap 
model\cite{DYRE:1995p2812,G64}, we will discuss the results of a temperature jump.
Guided by the fact noted above that a trap model can - at least qualitatively - be used to describe the relaxation of the MB-energies of simulated supercooled liquids, we will study the behavior of the IS- and MB-energies of the Kob-Andersen KABLJ liquid.
\section*{II. Energy relaxation in the Gaussian trap model}
In the trap model one considers a collection of traps, to be termed 'states' in the following, characterized by their energies.
The temporal evolution of the population of these states is determined by simple kinetic rules.
After the activated transition out of a given state, the destination trap/state is chosen at random, i.e. according to the prescribed density of states (DOS) $\rho(\e)$ of trap energies $\e$.
The escape rate for the transitions out of the state characterized by $\e$ is given by the Arrhenius law $\k_T(\e)=\k_\infty e^{\b\e}$ where $\k_\infty$ denotes the attempt frequency and 
$\b=(k_BT)^{-1}$ with the Boltzmann constant $k_B$.
The master equation for the populations is given in the Appendix for the convenience of the reader.

Different realizations of the trap model have been considered in the past.
Often, one considers the model with an exponential distribution of trap 
energies\cite{Monthus:1996p3207}.
This model shows a phase transition into a low-temperature phase where equilibrium can never be reached.
However, when discussing canonical glasses one usually deals with systems that reach equilibrium in the long run as in the Gaussian trap model\cite{Heuer:PETReview,DYRE:1995p2812,Denny:2003p3738,G56}.

In the following, we briefly review the results that one obtains for this model in order to motivate the MD-simulations presented afterwards.

\begin{figure}[!h]
\centering
\includegraphics[width=7.5cm]{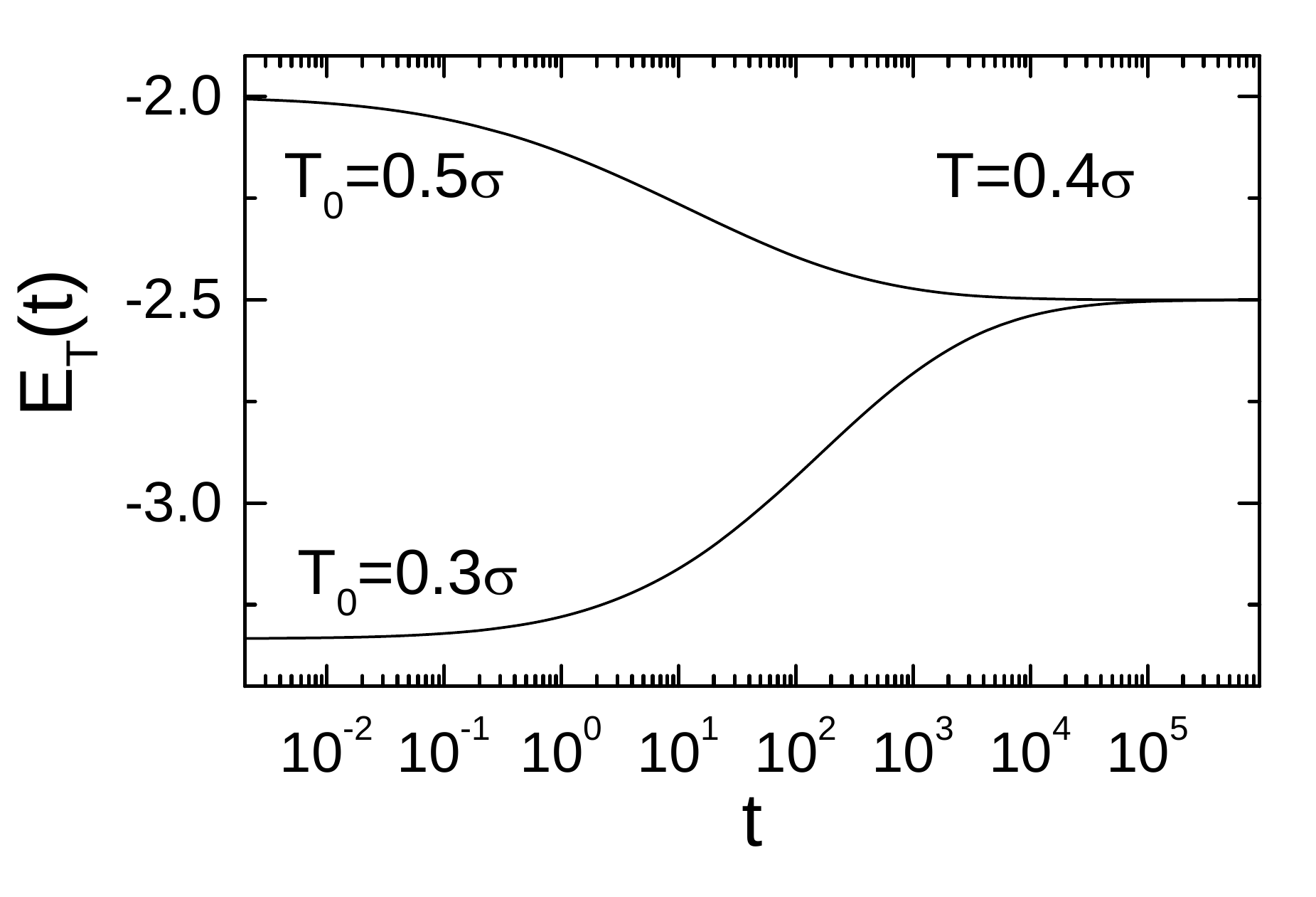}
\vspace{-0.5cm}
\caption{Temporal evolution of the mean energy $E_T(t)$ for a jump from $T_0=0.3\s$ to $T=0.4\s$ and for $T_0=0.5\s$ to $T=0.4\s$. $E_T(t)$ is calculated according to eq.(\ref{E.n.t}) using the numerical solution of the master equation, eq.(\ref{ME.p}). The time is measured in units of the attempt frequency $\k_\infty$.}
\label{Fig.1}
\end{figure}
We start with considering the relaxation of the energy after a sudden change from some initial temperature $T_0$ to a destination temperature $T$. 
In Fig.\ref{Fig.1} we show results for both, a quench and an up-jump.
It is obvious that the value of the mean energy, $E_T(t)$, changes monotonously from $E_{T_0}^{\rm eq}=-\b_0\s^2$ to $E_T^{\rm eq}=-\b\s^2$ in both cases.
If one compares the behavior for the quench and the up-jump, there is an asymmetry in the relaxation\cite{Angell:2000p3651}.
This asymmetry reflects the fact that linear-response theory cannot be applied for the interpretation of such large temperature jumps (in linear-response, the curves are symmetric).
The relaxation of the energy looks very similar to the curves shown in Fig.\ref{Fig.1} if other initial and final temperatures are considered.

\begin{figure}[!h]
\centering
\includegraphics[width=7.5cm]{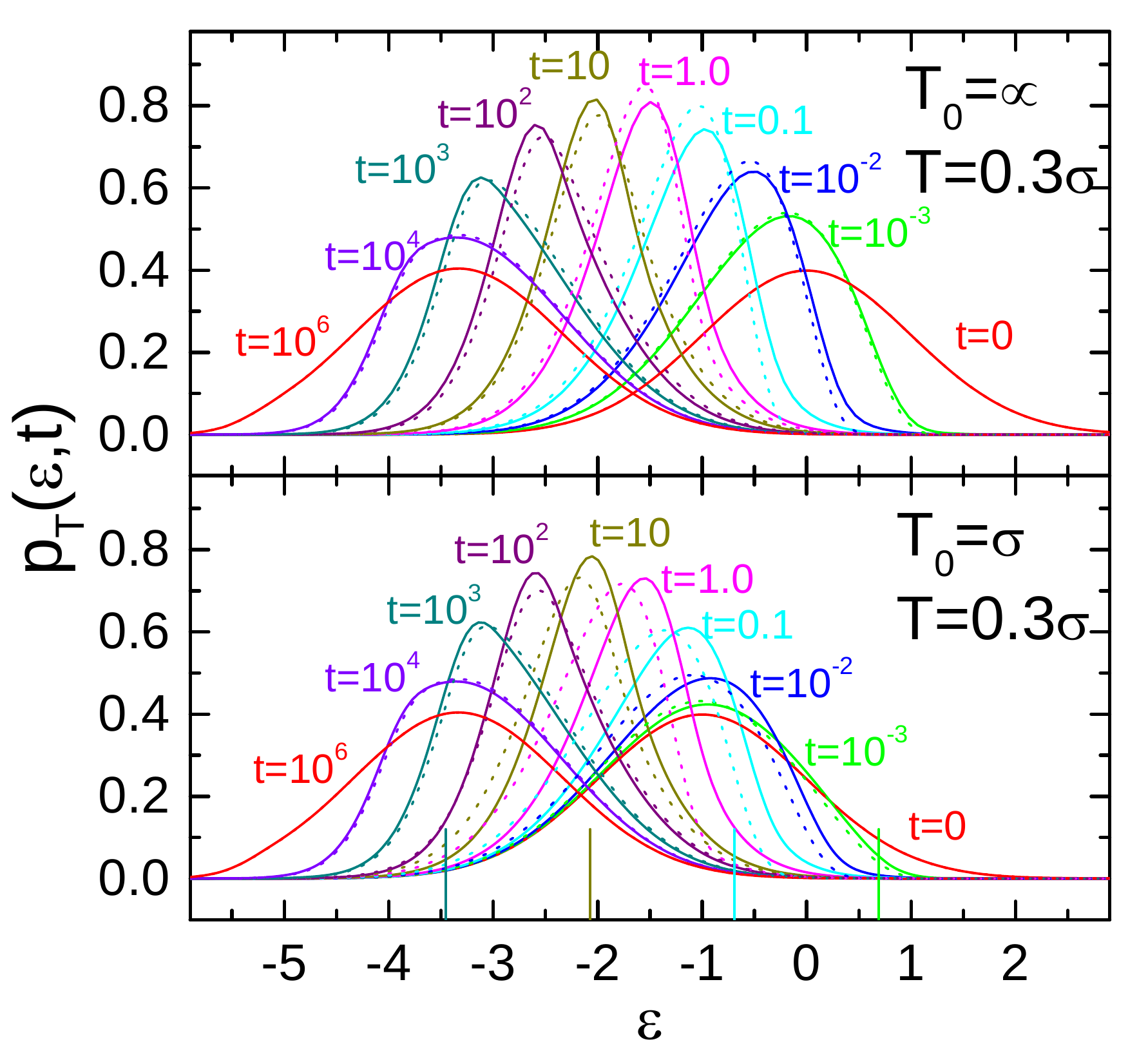}
\vspace{-0.5cm}
\caption{Trap populations $p_T(\e,t)$ for different times (measured in units of the attempt frequency $\k_\infty$) after a quench from $T_0=\infty$ (upper panel) and $T_0=\s$ (lower panel) to $T=0.3\s$.
The full lines represent the numerical solution of the master equation, eq.(\ref{ME.p}).
The dotted lines are obtained from the approximative solution given in eq.(\ref{p.app.t}) and the vertical lines in the lower panel denote the demarcation energy $\e_D$ for $t=10^{-3}$, $10^{-1}$, 
$10^{1}$, $10^{3}$ (from right to left)}
\label{Fig.2}
\end{figure}

In the present context the temporal evolution of the distributions of the energies, $p_T(\e,t)$, is more important than its moments.
Therefore, in the following we will discuss the behavior of $p_T(\e,t)$ after a temperature jump, cf.\cite{G56}. 
In Fig.\ref{Fig.2} we show the evolution of $p_T(\e,t)$ after a quench starting from two different initial temperatures $T_0$. 
It is evident immediately that the Gaussian is centered at $E_T(t=0)=-\b_0\s^2$ directly after the quench and at $E_T(t=\infty)=-\b\s^2$ after the system has relaxed to the final temperature.
The full lines correspond to the numerical solution of the master equation, while the dotted lines are based on the following approximation.
Neglecting any correlations in the energies of the inital and final trap of a transitions, one can assume that the initial distribution $p^{\rm eq}_{T_0}$ diminishes due to escape-transitions according to $p^{\rm eq}_{T_0}(\e)e^{-\k_T(\e)t}$.
Similarly, the distribution at the destination temperature, $p^{\rm eq}_T$, is built up independently
according to $p^{\rm eq}_T(\e)\left(1-e^{-\k_T(\e)t}\right)$.
These considerations allow us to write:
\be\label{p.app.t}
p^{\rm ap.}_T(\e,t)=N(t)^{-1}\left[ p^{\rm eq}_{T_0}(\e)e^{-\k_T(\e)t}
                                   +p^{\rm eq}_T(\e)\left(1-e^{-\k_T(\e)t}\right) \right]
\ee
where
\[
N(t)=1+\int\!d\e\left[p^{\rm eq}_{T_0}(\e)-p^{\rm eq}_T(\e)\right]e^{-\k_T(\e)t}
\]
is a constant that ensures the normalization of $p^{\rm ap.}_T(\e,t)$.
Actually, one can show that eq.(\ref{p.app.t}) follows from the long-time limit of the solution of the master equation, eq.(\ref{ME.p}).
It is intriguing how well this simple approximation represents the numerical solution of the master equation.
The quality of the approximation (\ref{p.app.t}) is similar for other $T_0$ and $T$.
Eq.(\ref{p.app.t}) can be further interpreted in the following way. 
At a given time the states can be divided into two classes, those that have undergone an escape-transition, i.e. those that have relaxed, and those that are still frozen.
The latter are low-energy states with $e^{-\k_T(\e)t}\sim1$ and for the former one has 
$e^{-\k_T(\e)t}\sim0$.
This distinction between frozen and relaxed states can be quantified via the definition of the 
so-called demarcation energy via $\k_T(\e_D)=1/t$, i.e.
\be\label{E.dem}
\e_D(t)=-\b^{-1}\ln{\left(\k_\infty t\right)}
\ee
The meaning of $\e_D$ is that, due to the activated nature of the dynamics, most states with 
$\e\!<\!\e_D$ are frozen at time $t$, while those with $\e\!>\!\e_D$ essentially have reached 
equilibrium\cite{DYRE:1995p2812,DYRE:1987p2799,Arkhipov1979}.
Eq.(\ref{p.app.t}) also allows a simple interpretation of the temporal evolution of the moments of the distribution.
For a given time states with $\e<\e_D$ are still represented by $p^{\rm eq}_{T_0}(\e)$ while those with larger energy have already relaxed and are represented already by $p^{\rm eq}_{T}(\e)$. 
The different weights of the initial and final Gaussian give rise to a narrowing of the distribution at intermediate times, as observed in the quoted simulations\cite{Sciortino2004}.

\begin{figure}[!h]
\centering
\includegraphics[width=7.5cm]{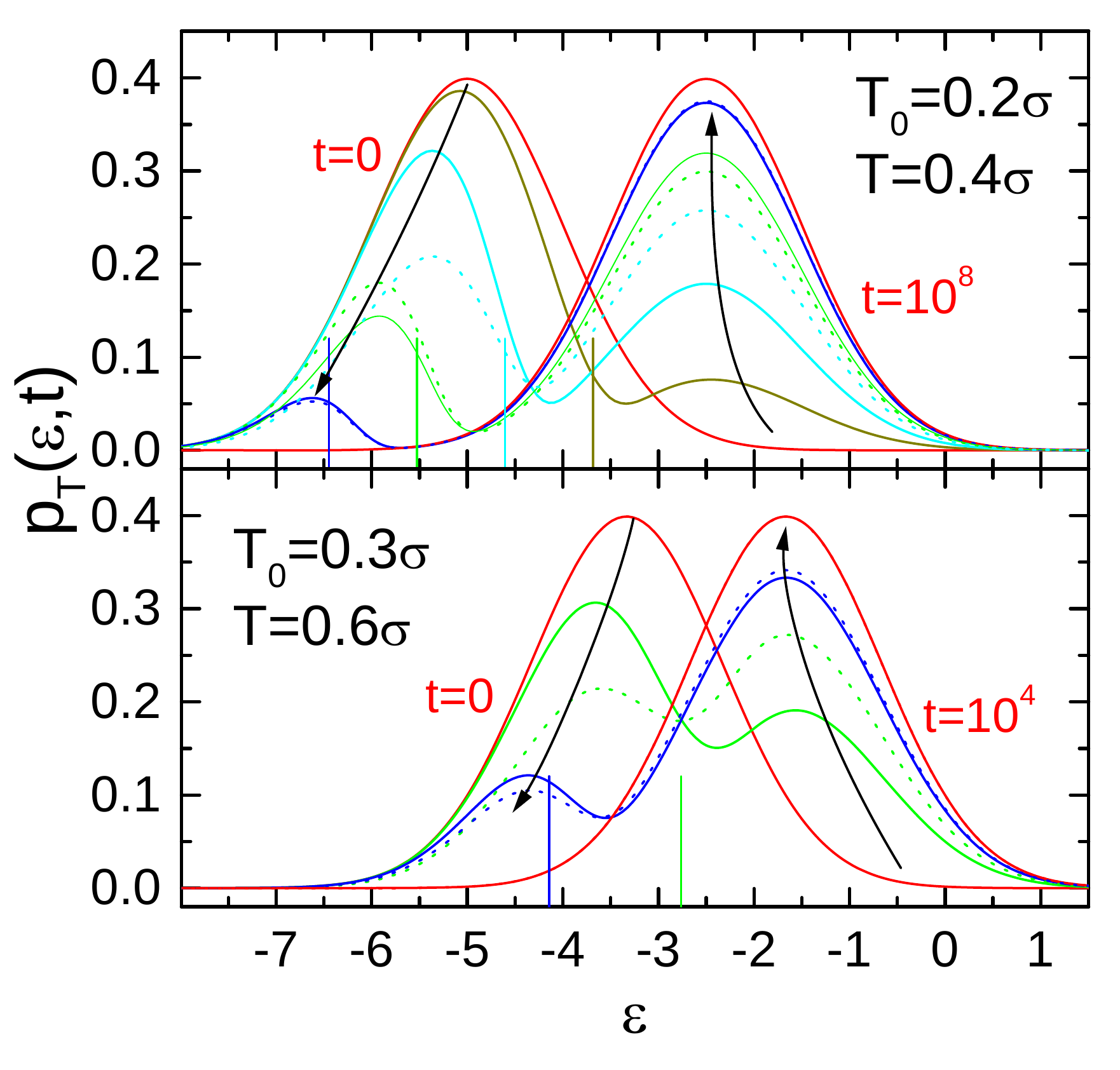}
\vspace{-0.5cm}
\caption{$p_T(\e,t)$ for different times after a temperature up-jump from $T_0=0.2\s$ to $T=0.4\s$ (upper panel; $t=0, 10^2, 10^3, 10^4, 10^5, 10^8$s) and from $T_0=0.3\s$ to $T=0.6\s$ (lower panel; $t=0, 1, 10, 10^4$s).
The solid lines are the numerical solution of the master equation and the dotted lines represent the approximation (\ref{p.app.t}).
The vertical lines show the demarcation energies for the various times.
Note that with increasing time $\e_D$ decreases meaning that less states are frozen for longer times.}
\label{Fig.3}
\end{figure}

The situation changes completely if instead of a quench from a high to a low temperature the opposite case of an up-jump from $T_0<T$ is considered, cf. Fig.\ref{Fig.3}.
In this case, $p^{\rm eq}_{T_0}(\e)$ is centered at lower energy than $p^{\rm eq}_{T}(\e)$ which means that it is the slowly relaxing part of the distribution that dominates the dynamics.
Put differently, the high energy tail of $p^{\rm eq}_{T_0}(\e)$ relaxes quickly and thus 'moves' to higher energies, while the low energy part is trapped for a very long time.
Thus, one has a distribution that can be viewed as as a weighted superposition of the two involved equilibrium distributions, separated roughly by the demarcation energy, cf. the vertical lines in Fig.\ref{Fig.3}.
This scenario gives rise to a two-peak structure, cf. Fig.\ref{Fig.3}, that was discussed earlier by one of us\cite{DYRE:1995p2812}. 
The dotted lines are the approximate solution, eq.(\ref{p.app.t}). 
It is evident that the quality of the approximation increases with increasing time.
Also the behavior of the moments of $p_T(\e,t)$ for an up-jump is evident from Fig.\ref{Fig.3}. 
The two-peak behavior already indicates that here the second moment will show a maximum at some intermediate time.

\begin{figure}[!h]
\centering
\includegraphics[width=7.5cm]{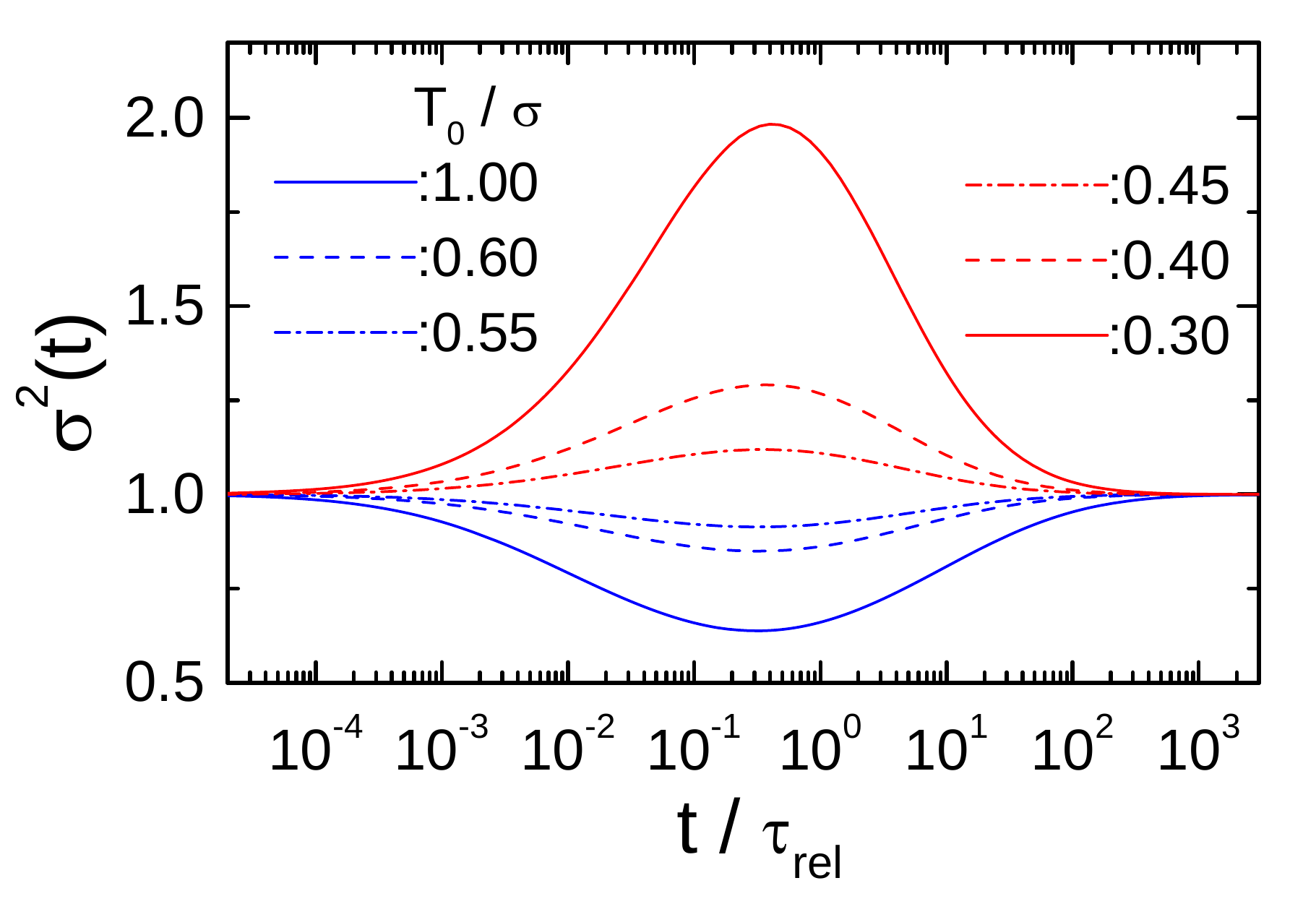}
\vspace{-0.5cm}
\caption{$\s^2(t)$ for a temperature jump with different initial temperatures $T_0$ to 
$T=0.5\s$ versus time scaled by the temperature-dependent relaxation time 
$\t_{\rm rel}$; quench: $T_0/\s=1.0$, $0.6$, $0.55$ (blue) and up-jump: $T_0/\s=0.3$, $0.4$, $0.45$ (red)}
\label{Fig.4}
\end{figure}

In Fig.\ref{Fig.4} we show the variance of $p_T(\e,t)$ for both, various quenches and up-jumps.
We scaled the time by the relaxation time $\t_{\rm rel}$.
This relaxation time is determined from the temporal evolution of $E_T(t)$, cf. Fig.\ref{Fig.1}, as the $1/e$-decay time of the normalized energy-relaxation function
$(E_T(t)-E_T^{\rm eq})/(E_{T_0}^{\rm eq}-E_T^{\rm eq})$ and roughly divides the energies into those that have and those that have not relaxed until $t=\t_{\rm rel}$.
Accordingly, the maximum/minimum width is found for $t\sim\t_{\rm rel}$ (more precisely around 
$t/\t_{\rm rel}\sim 0.3\cdots 0.4$).

Given the fact that the equilibrium behavior of simulated glass-forming liquids and the aging dynamics following a quench can qualitatively be understood in terms of a Gaussian trap model we anticipate that it should be possible to observe the two-peak structure in the distributions following an up-jump also in simulations.
\section*{III. Energy relaxation in the KABLJ liquid}
We now discuss the results of molecular dynamics simulations of the KABLJ liquid\cite{KA1995a,KA1995b}.
We start with a jump from high to low temperatures very similar to what has already been considered by 
Saika-Vovoid and Sciortino\cite{Sciortino2004}, and afterwards we will show that one indeed can observe a double-peak structure when a jump from a low to a high temperature is considered.

The simulations of what will be called the {\it large system} have been carried out on a KABLJ system of 1000 particles. 
All simulation were performed in the NVT ensemble, and time and temperatures are in reduced units (the density is 1.204). 
Temperature was controlled using a Nose-Hoover thermostat and a time step of 0.001 was used.
The distributions of the inherent state energies were obtained by averaging over 1500 independent runs (each of length 4000) both for the up-jumps and the quenches. 

In order to investigate the distributions of metabasin (MB) energies, we additionally performed simulations on a {\it small system}.
We use a KABLJ system of 65 particles (and a reduced cut-off radius $r_c=1.8$) in the NVT ensemble at a density of 1.2.
We determined the distribution of MB-energies as described in detail in ref.\cite{MBHeuer2000}.
For the up-jump we used 4600 trajectories of length 10000 and for the quench we used 3000 trajectories of length 70000.
The LJ-system with the shorter cutoff has slightly different properties compared to using the larger cutoff. However, the modifications are small ($T_c$, e.g., is roughly modified by 2\%). 
Since we do not compare absolute values of energies, this modification is of little relevance for the present work.

We have checked\cite{Denny:2003p3738} that the distribution of MB and IS is nearly identical in the temperature range of interest for this work. 
This is not surprising, because during the stay in a given MB the system is mainly residing in the IS with the lowest energy. 
Thus, considerable differences between IS and MB only come into play when analyzing the dynamics (see below).
\subsection*{1. Large system}
In Fig.\ref{Fig.5} we present results from simulations of a quench from $T_0=0.6$ to $T=0.45$.
\begin{figure}[!h]
\centering
\includegraphics[width=7.5cm]{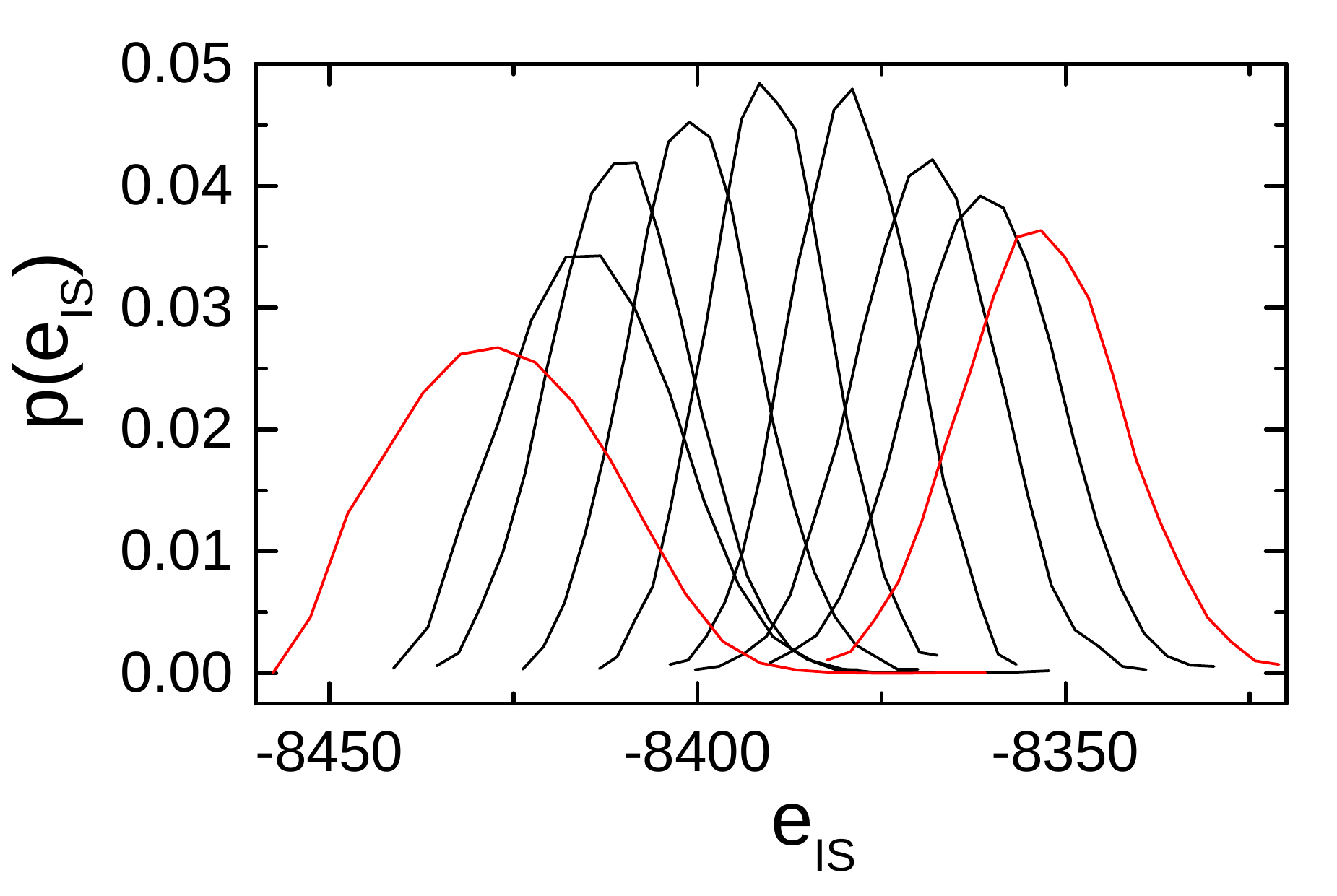}
\vspace{-0.5cm}
\caption{$p(e_{IS})$ or the large system for different waiting times, t=9.6, 27.0, 76.8, 614.4, 868.8, 1737.6, 3475.8 from right to left after a quench from $T_0=0.6$ to $T=0.45$ (black lines).
The red lines are the equilibrium distributions at the initial and the final temperature.}
\label{Fig.5}
\end{figure}
The results are very similar to those of Saika-Vovoid and Sciortino\cite{Sciortino2004}.
The distribution moves from high energies at high temperature to lower energies as a function of the time elapsed after the quench, and the width of the distribution diminishes for intermediate times.
The overall behavior is very similar to what is observed in the trap model, cf. Fig.\ref{Fig.2}, with the difference that the widths of the equilibrium distributions are temperature-dependent as opposed to the situation in the Gaussian trap model.
\begin{figure}[!h]
\centering
\includegraphics[width=7.5cm]{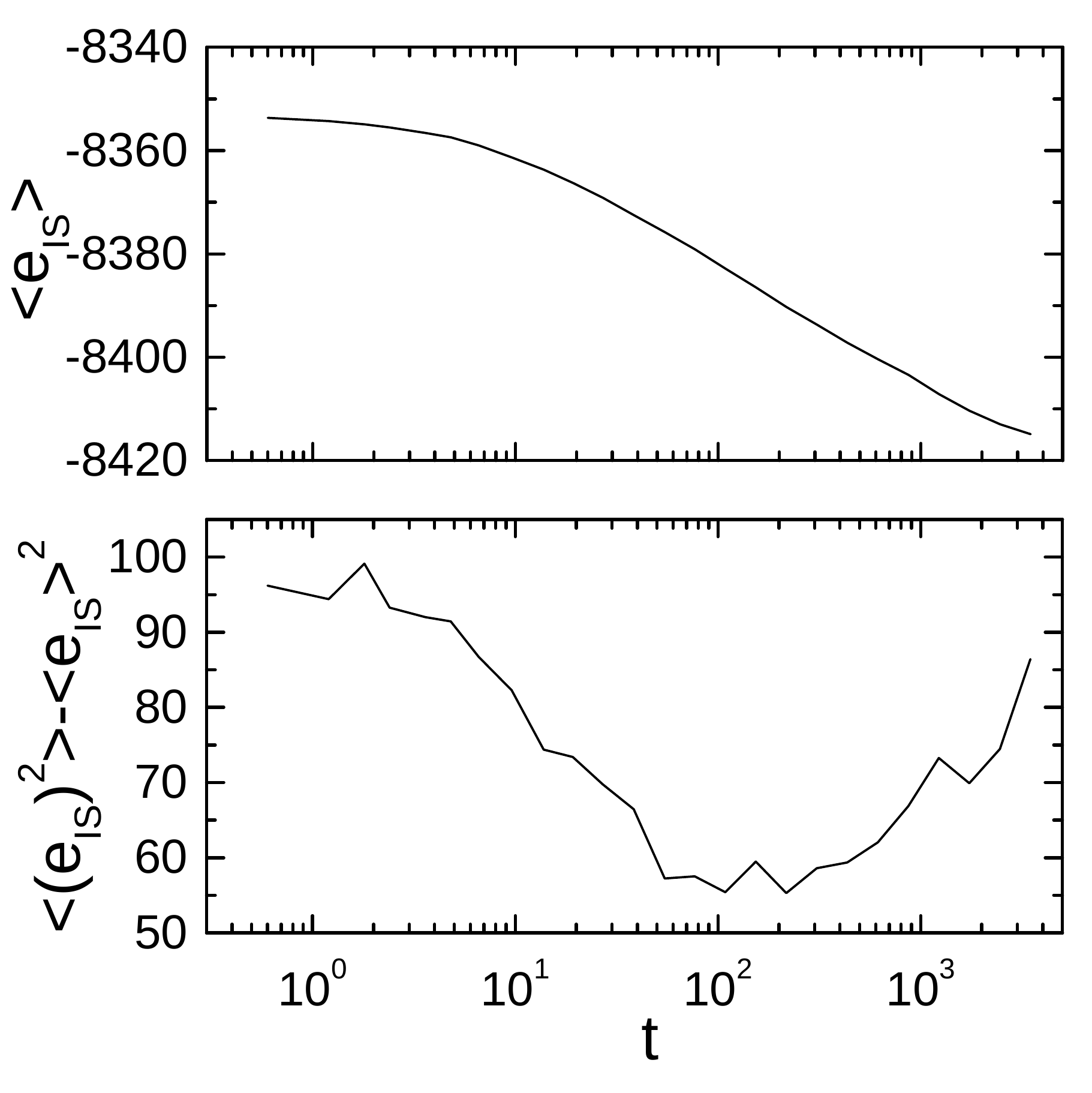}
\vspace{-0.5cm}
\caption{First and second moment of $p(e_{IS})$ for the large system as obtained from MD simulations after a quench from $T_0=0.6$ to $T=0.45$, for details see the main text.}
\label{Fig.6}
\end{figure}

The moments of the distribution are shown in Fig.\ref{Fig.6}. 
In particular the monotonous decay of the mean and the drop in the variance for intermediate times are clearly displayed. 

We have seen that the simulation data for a quench can qualitatively be understood in terms of a Gaussian trap model.
From this model we expect a bimodal distribution to show up for intermediate waiting times, reflecting the different relaxation behavior of states below and above the demarcation energy if an up-jump in temperature is considered instead of a quench.

\begin{figure}[!h]
\centering
\includegraphics[width=7.5cm]{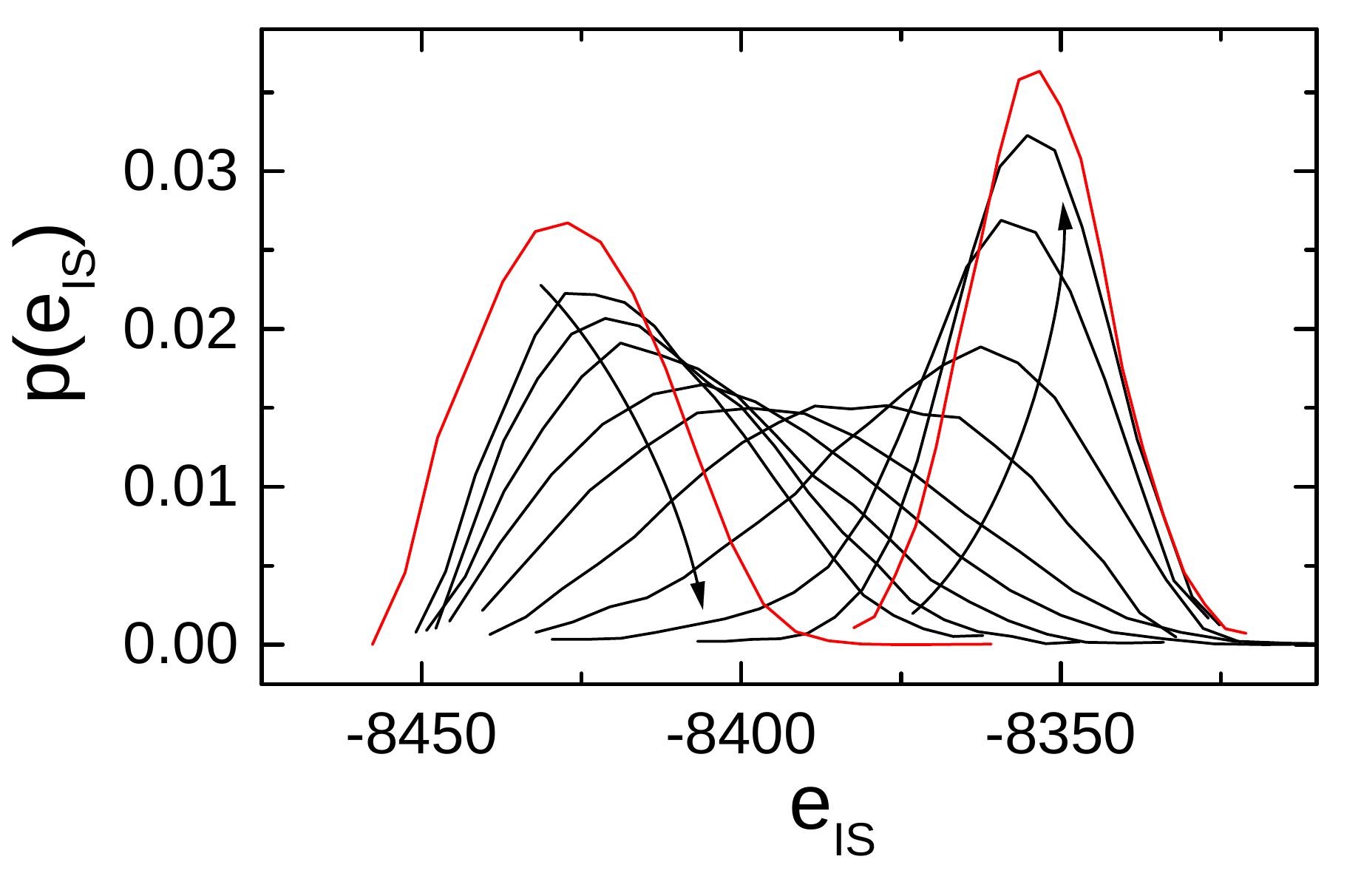}
\vspace{-0.5cm}
\caption{$p(e_{IS})$ for the large system for different waiting times, t=27.0, 38.4, 54.6, 76.8, 108.6, 153.6, 217.2, 307.2, 434.2 (black lines, in the order indicated by the arrows) after a temperature jump from $T_0=0.45$ to $T=0.6$. The red lines correspond to the equilibrium distributions at the respective temperatures.}
\label{Fig.7}
\end{figure}

In Fig.\ref{Fig.7} we present $p(e_{IS})$ for an up-jump from $T_0=0.45$ to $T=0.6$. 
Even though one cannot observe a bimodal distribution it is obvious that the distribution broadens tremendously for intermediate times, cf. the curves for t=108.5 and t=153.6.
Low-energy states relax extremely slowly, while states with higher IS-energy move towards the equilibrium distribution at the higher temperature much more rapidly.
This means that for short times only a few (high-energy) states are relaxed, while for long times there are still some low-energy states that are not yet relaxed.

\begin{figure}[!h]
\centering
\includegraphics[width=7.5cm]{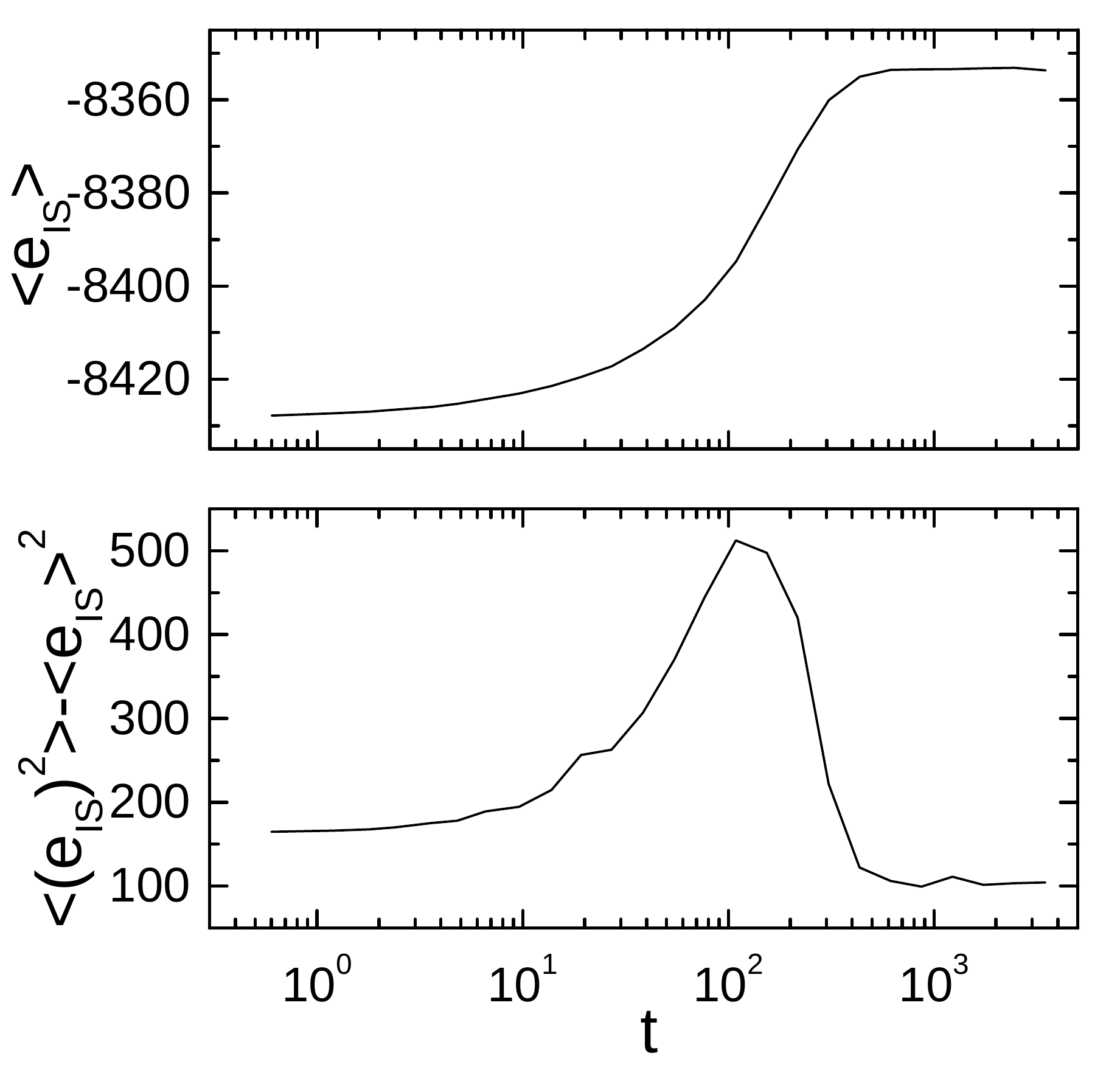}
\vspace{-0.5cm}
\caption{First and second moments of $p(e_{IS})$ for the large system for an up-jump from $T_0=0.45$ to $T=0.6$}
\label{Fig.8}
\end{figure}

The corresponding moments are shown in Fig.\ref{Fig.8}. 
These curves nicely reflect the fact that the variance reaches its maximum value roughly when the time elapsed after the up-jump coincides with the relaxation time of $E_T(t)$, indicating that about half of the inherent structures are relaxed at this time while the others still are frozen at their low-temperature values.
These simulations indicate that the predictions of the trap model might be helpful in the interpretation not only of the equilibrium properties but also the aging behavior of the system studied.
However, for the up-jump we do not observe a two-peak structure of the distributions of IS-energies.
The reason will be outlined below.
\subsection*{2. Small system}
In the following we present simulation results for the small systems. The system size of 65 particles is small enough to allow the identification of the MBs and we will concentrate on the temporal evolution of the distribution of MB-energies, $p(e_{MB})$, after sudden temperature changes.

\begin{figure}[!h]
\centering
\includegraphics[width=7.5cm]{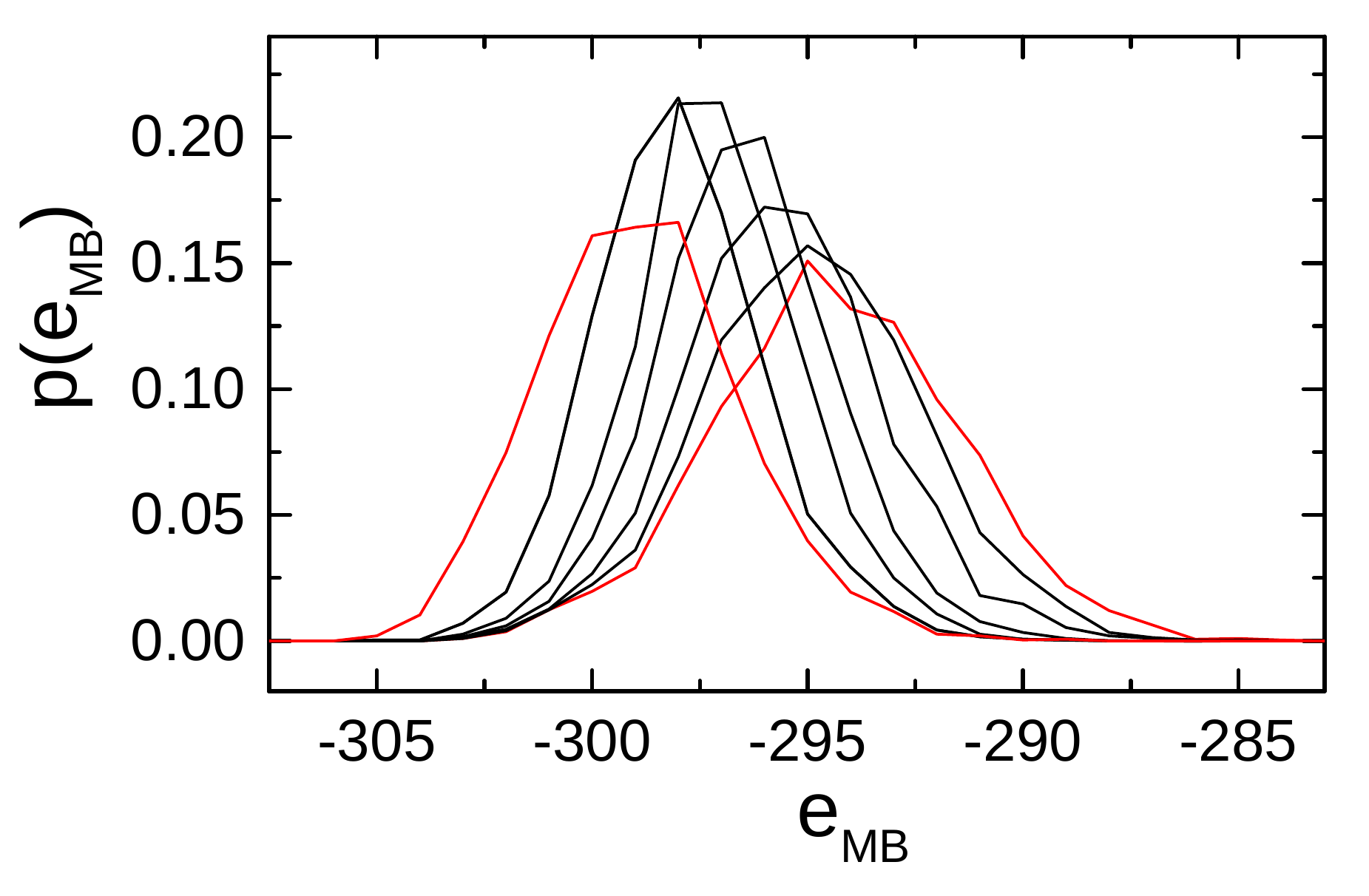}
\vspace{-0.5cm}
\caption{$p(e_{MB})$ of the small system for different waiting times, t=4, 32, 152, 392, 1992, 15992 (from right to left) after a quench from $T_0=0.6$ to $T=0.45$ (black lines).
Red lines: $p^{\rm eq}_T(e_{MB})$}
\label{Fig.9}
\end{figure}

In order to compare the behavior of the IS-energies to the corresponding behavior of the MB-energies, we show the evolution of $p(e_{MB})$ for the same quench from $T_0=0.6$ to $T=0.45$ for the small system in
Fig.\ref{Fig.9}.
Apart from the smaller values of the MB-energies the overall features like the shift of the distribution and the narrowing at intermediate times are very similar to the observations made for the large system in Fig.\ref{Fig.5}.
\begin{figure}[!h]
\centering
\includegraphics[width=7.5cm]{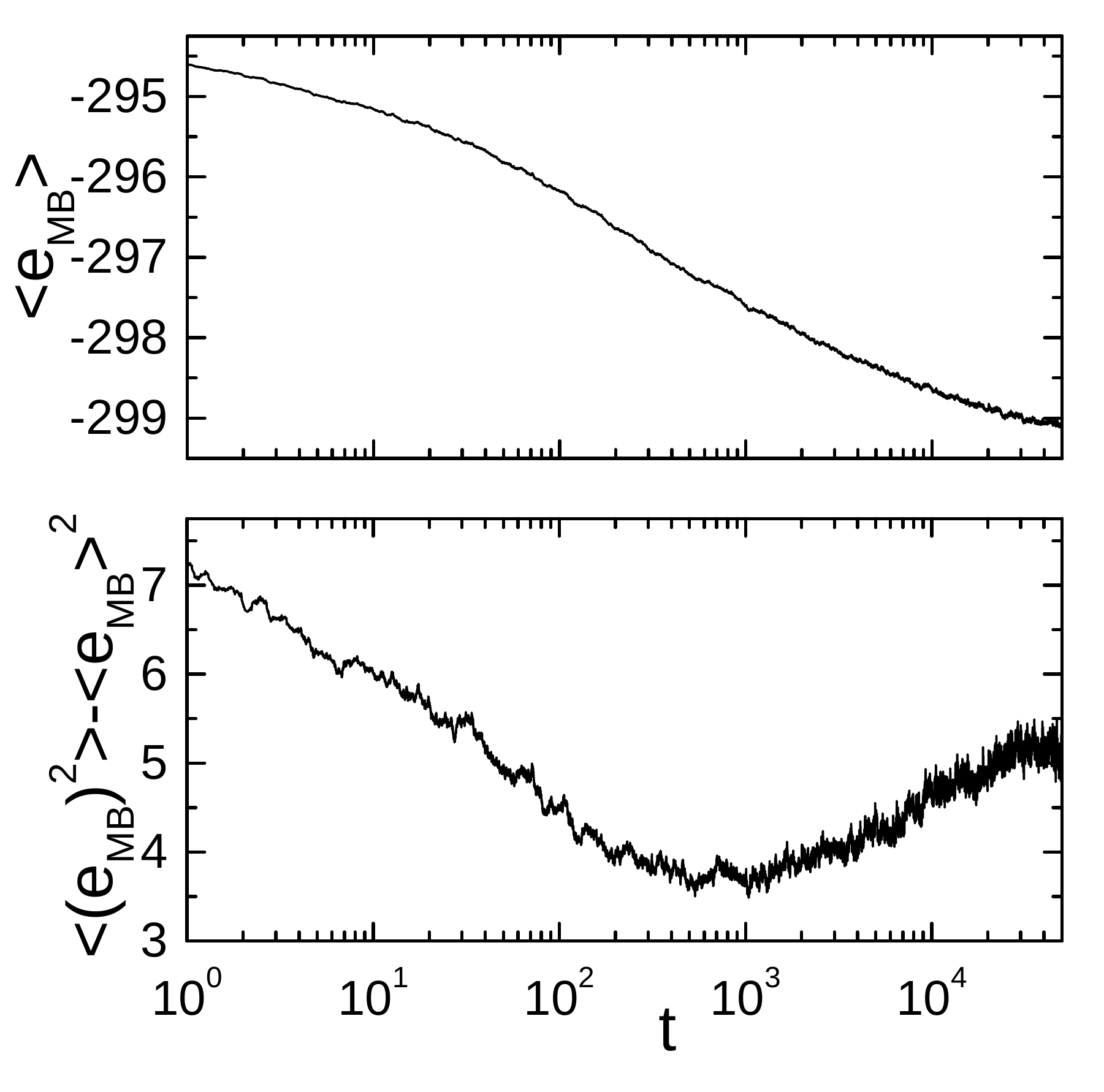}
\vspace{-0.5cm}
\caption{First and second moment of $p(e_{MB})$ for the small for a quench from $T_0=0.6$ to $T=0.45$.}
\label{Fig.10}
\end{figure}
In Fig.\ref{Fig.10} we plot the moments of the distribution as a function of the waiting time.
This plot shows that also the moments of $p(e_{MB})$ behave very much in the same way as those of the distributions of the IS-energies in the large system. 
The most prominent differences are the better statistics and the increase in the overall relaxation time of roughly one decade in case of the small system. 

We have seen that the evolution of both, the IS-energies or the MB-energies again qualitatively follow the predictions of the trap model with a Gaussian DOS.
Thus, the dependence of the moments on the waiting time can be understood in terms of the demarcation energy, which separates low-energy states that up to the given time are frozen from high-energy states, which at the same time have reached the new equilibrium already.

\begin{figure}[!h]
\centering
\includegraphics[width=7.5cm]{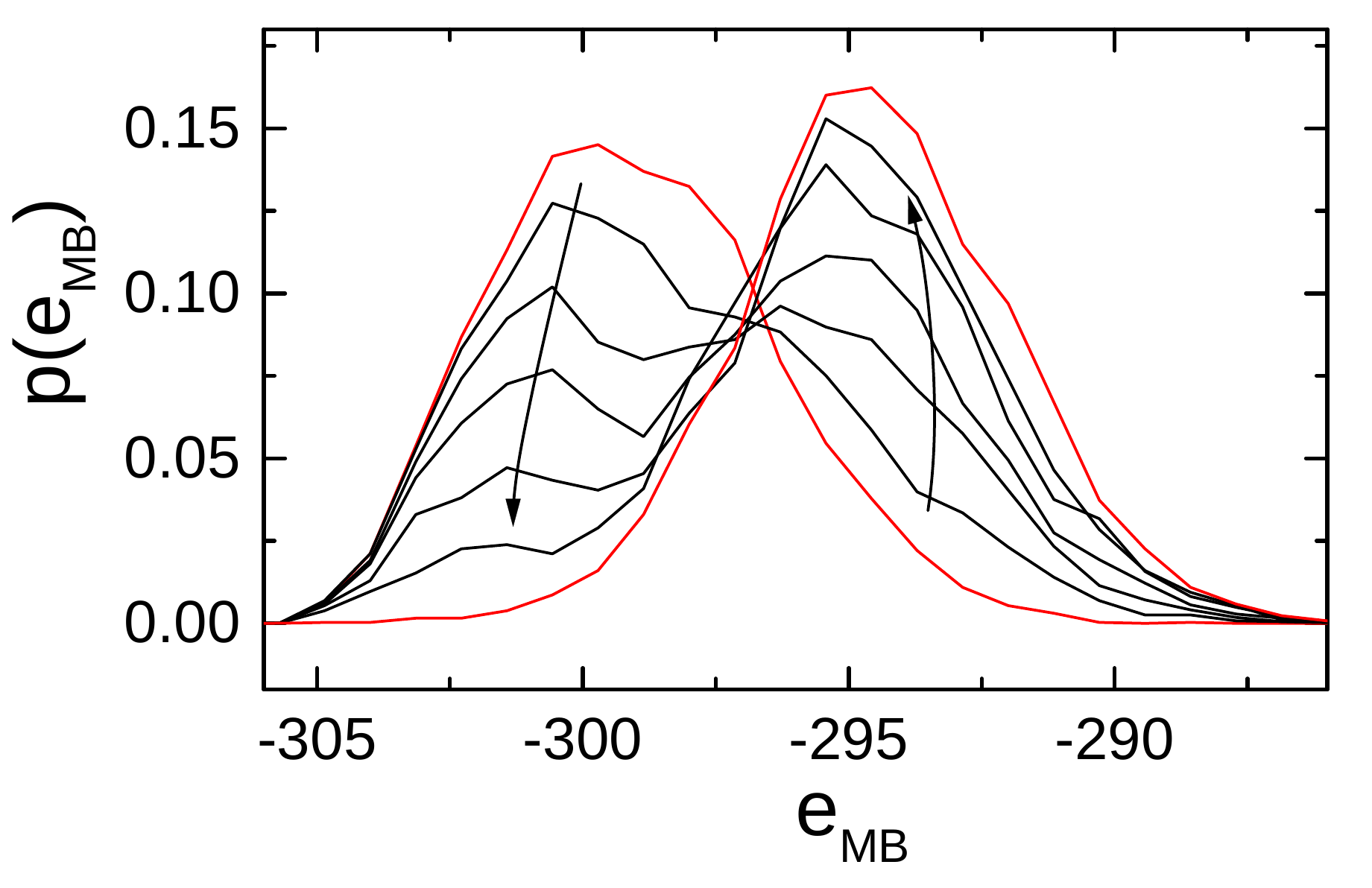}
\vspace{-0.5cm}
\caption{$p(e_{MB})$ for the small system for different waiting times, t=115, 240, 390, 740, 1490 
after an up-jump from $T_0=0.45$ to $T=0.6$ (black lines, in the order indicated by the arrows).
Red lines: $p^{\rm eq}_T(e_{MB})$}
\label{Fig.11}
\end{figure}

\begin{figure}[!h]
\centering
\includegraphics[width=7.5cm]{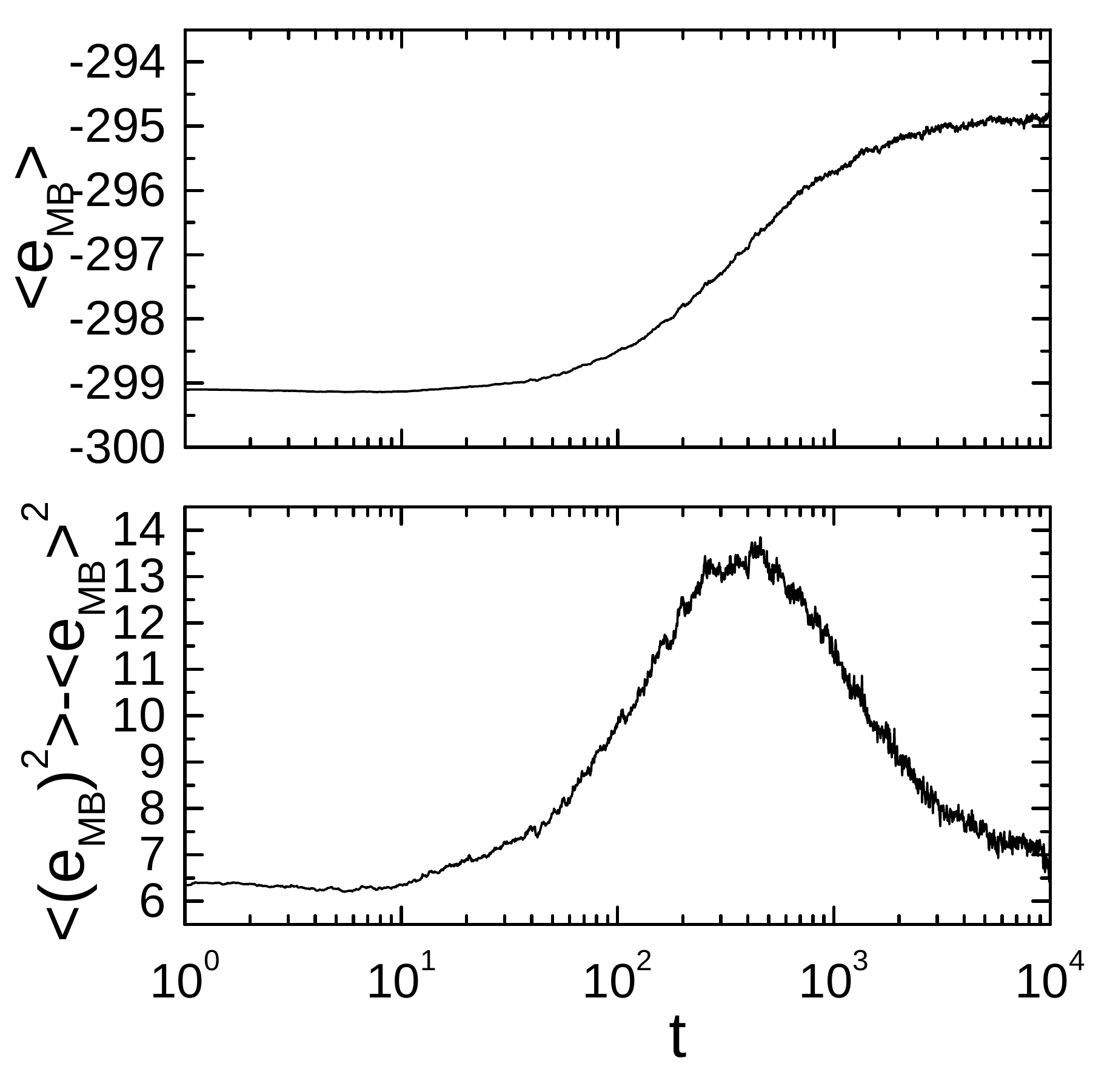}
\vspace{-0.5cm}
\caption{First and second moment of $p(e_{IS})$ for an up-jump from $T_0=0.45$ to $T=0.6$}
\label{Fig.12}
\end{figure}

Next, we consider the same up-jump from $T_0=0.45$ to $T=0.6$ as for the large system.
We present the distributions $p(e_{MB})$ for various times after the up-jump in Fig.\ref{Fig.11}.
In this case we really observe a two-peak structure in the distribution for intermediate waiting times.
This effect is most prominent for times on the order of the relaxation time, i.e. on the order of $300\cdots400$, cf. Fig.\ref{Fig.12}.

\begin{figure}[!h]
\centering
\includegraphics[width=7.5cm]{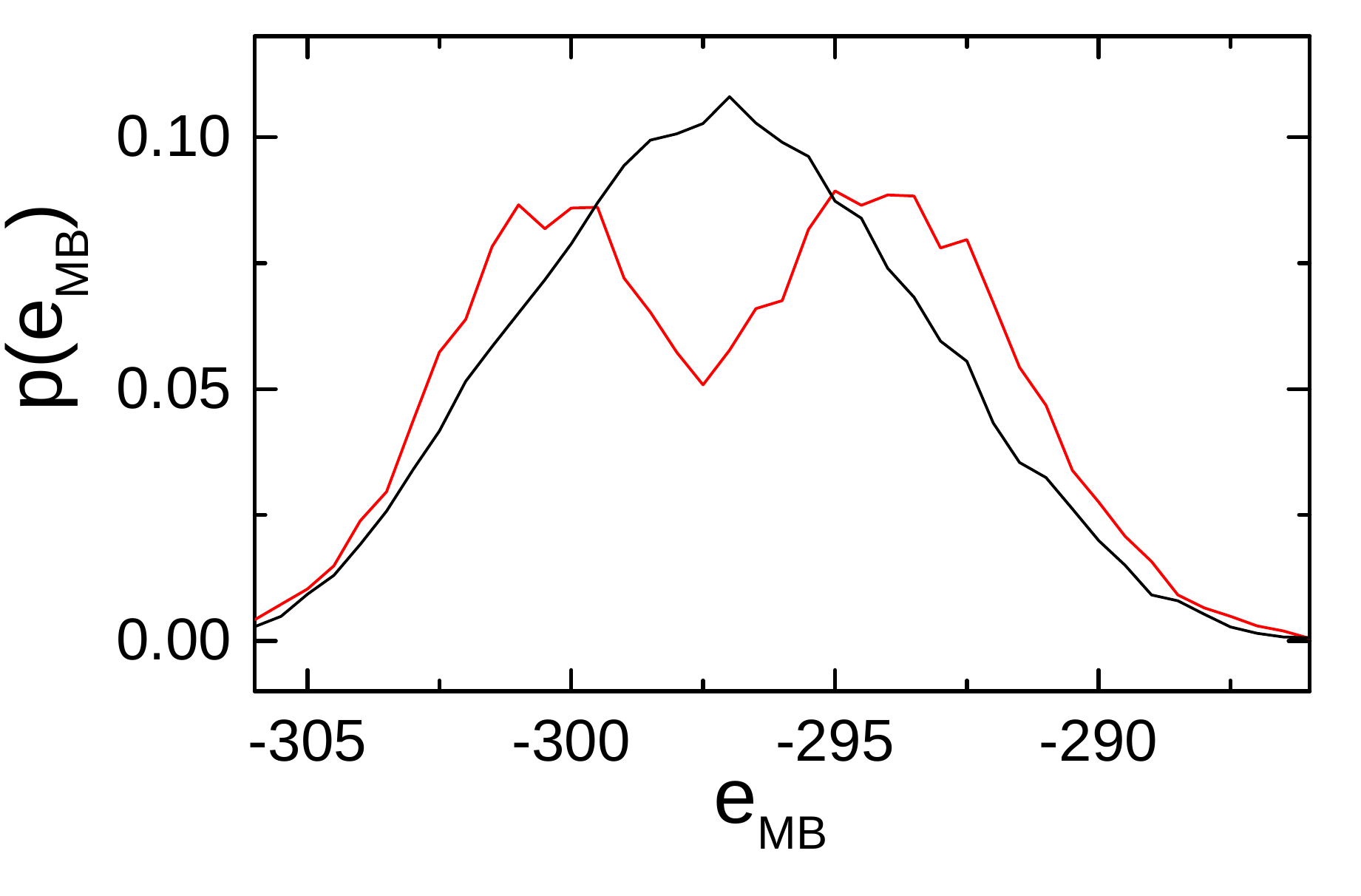}
\vspace{-0.5cm}
\caption{$p(e)$ for the extended trap model with $N=1$ (red line) and $N=2$ (black line) small systems after a temperatur jump from $T=0.45$ to $T=0.6$. The waiting times correspond to the time when the variance reaches its maximum value.}
\label{Fig.13}
\end{figure}

One may wonder why the small system displays a two-peak structure but not the large system.
To clarify this point we compare the case of a single trap model (N=1) with the case of two superimposed trap models (N=2), see Fig.\ref{Fig.13}.
To render the superposition more realistic we have adopted the dynamic coupling as described in 
ref.\cite{HDS2005}.
As shown in that reference, a LJ system of 65 particles should indeed be consideredas a superposition of 2-3 elementary trap models, reflecting the fact that the elementary relaxation processes, i.e. MB transitions, are on average localized to 20-30 particles.
Interestingly, explicit simulations of the N=2 case does not yield a two-peak structure, cf. Fig.\ref{Fig.13}.
Thus, it is not at all surprising that also the much larger system discussed above does not display this feature.
One should rather wonder why the real 65-particle system displays a two-peak structure.
Note that it is less pronounced than for the N=1 case shown in Fig.\ref{Fig.13}.
A possible reason is as follows.
The degree of localization of the MB transition can vary from case to case.
Thus, for some transitions the total system may be described as one elementary system, i.e. N=1, whereas in other cases larger values of N may characterize the present configuration.
The average over the different situations, i.e. an average over a two-peak and a single-peak distribution, can yield a two-peak distribution albeit with a less pronounced separation between both peaks.
Stated differently, we interprete the emergence of the two peaks as a consequence of the presence of different spatial extensions for different MB transitions.
From this observation we conclude that for a small system as considered here couplings among different 'regions' are of minor relevance only and that a description of the aging dynamics in terms of a Gaussian trap model captures the most important dynamic features.

\begin{figure}[!h]
\centering
\includegraphics[width=7.5cm]{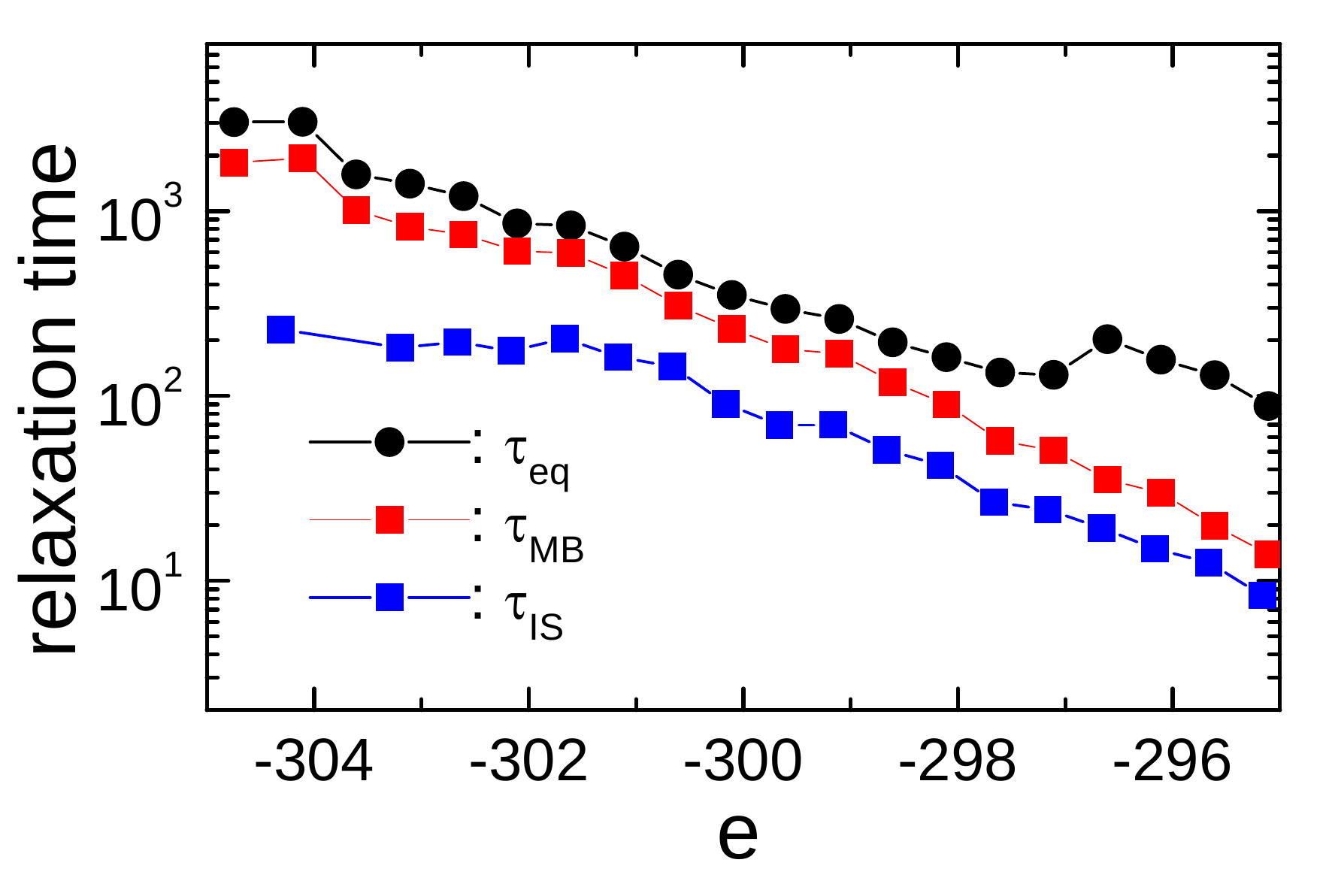}
\vspace{-0.5cm}
\caption{Time scale of energy relaxations $\t_{\rm eq}$ vs. escape time $\t_{IS}$ and $\t_{MB}$ after a temperatur jump from $T=0.45$ to $T=0.6$; see text for the definition of the various relaxation times.
}
\label{Fig.14}
\end{figure}

A more quantitative comparison with the trap model is possible by comparing the time scale of energy equilibration $\t_{\rm eq}$ after a temperature jump with the time scale to escape an IS or an MB 
($\tau_{IS}$ or $\tau_{MB}$, respectively) in their dependence of the initial energy. 
For the trap model a single jump yields full decorrelation of the energy. 
As shown in Fig.\ref{Fig.14} for the whole energy range one roughly has
$\t_{\rm eq}(e) \approx 2 \tau_{MB}(e)$. 
At first glance this result contradicts the trap model predictions, but can qualitatively be
explained in the following way. 
As explained in ref.\cite{HDS2005} the reason for the remaining deviations may be
related to the fact that even the small system with just 65 particles
has to be regarded as a superposition of two subsystems. 
Thus, a single relaxation process does not necessarily decorrelate the whole system.
As a result one gets $\t_{\rm eq}>\tau_{MB}(e)$.
Actually, the more quantitative analysis in ref.\cite{HDS2005} shows that roughly 10 MB transitions are necessary for energy equilibration.
How does this agree with the behavior of the relaxation times shown in Fig.\ref{Fig.14}?
For low initial energies the energy attained after a single transition on average is already significantly higher than the starting value.
Thus, the second jump occurs much faster than the initial one.
The same behavior is found for the following MB transitions and as a consequence the resulting MB transitions take place on a time scale $\tau_{MB}(e)$, giving rise to 
$\t_{\rm eq}/\t_{MB}(e)\approx2$.
For $e>-296$, however, this statement is no longer true because the first jump already is relatively fast.
Here, the complete factor of roughly 10 shows up in the ratio $\t_{\rm eq}/\t_{MB}(e)$.

In contrast, comparison of $\t_{\rm eq}(e)$ with $\tau_{IS}(e)$ clearly shows that for low energies
the deviations are as large as an order of magnitude.
This observation reflects the fact that at low temperatures and low IS energies the system displays many forward-backward processes between different IS which do not contribute to the energy relaxation. 
It is for this reason that the MB trajectories are better suited for a direct comparison with the trap model.
\section*{IV. Conclusions}
We presented molecular dynamics simulations of the temporal evolution of the distribution of inherent structure energies (of a large system) and the metabasin energies (of a small system) after sudden temperature changes in a model glass-forming liquid. 
While simulations for a quench from a high temperature to some low temperature exist already, we present new results for an up-jump from low to high temperature.

Both sets of results can qualitatively be understood in terms of the relaxation properties of the trap energies in a Gaussian trap model. 
In particular, the concept of a demarcation energy separating frozen states (lower energies) from states that are already relaxed (higher energies) is extremely helpful in rationalizing the data.
It allows to explain the behavior of the second moment in rather general terms.

Our most important result is that for an up-jump we find a two-peak structure in the distribution of MB-energies in case of the small system for times on the order of the relaxation time. 
This finding shows that the concept of the demarcation energy appears to qualitatively reflect the real behavior of the simulated system.
When the IS-energies in a large system are considered, one already sees the trend towards this bimodality in terms of extremely broad distributions. 
As discussed above, the fact that a large system is to be viewed as consisting of a number of (weakly interacting) sub-systems is responsible for the smearing of the effect that is so nicely visible in the small system.
This averaging of course is to be expected on the basis of the central limit theorem.

Of course, the trap model is a very simplistic mean-field model of the relaxation in glassy systems.
From our simulation results, however, we can speculate that a behavior similar to the one observed here might be rather generic for any Markovian model of collective energy jumps in systems with many degrees of freedom whenever a random walk description of the dynamics appears appropriate, i.e. if the correlations among the different energies decay fast.
Actually, recently the behavior of a large system has been described as a superposition of dynamically coupled elementary systems\cite{Rehwald2010}, i.e. a collection of trap models (using the present notation).
The present results suggest that a similar description might also work in non-equilibrium situations.

An asymmetry akin to the one discussed above between cooling and heating has been observed in experiments. In 1989 Bokov and Andreev\cite{Bokov1989} reported anomalously high intensity of (static) light scattering (at 90 degrees) from vitreous B$_2$O$_3$ during heating, but only at intermediate times when the system goes through the glass transition. Thus the previous cooling that produced the borate glass produced no such anomalous signal. Since the static light scattering intensity is proportional to the density fluctuations, the implication is that a glass upon heating temporarily goes through a phase characterized by anomalously large density fluctuations. Moynihan, Schroeder, and coworkers\cite{Moynihan1993, Schroeder1996} pointed out that this behavior is consistent with the presence of nanoscale density fluctuations which relax at different rates. If, as suggested by Brawer\cite{Brawer1984}, energy and density correlate locally such that low density corresponds to high energy, the trap model provides a simple way to understand these 
phenomena\cite{DYRE:1995p2812}. In this "volume-energy correlation" model the anomalous light scattering signals are direct proof of the above discussed two-peak structure in the energy distribution upon heating. 

If we assume that the energies considered in our present study (either IS- or MB-energies) can directly be related to activation energies relevant for the dynamics of supercooled liquids, we can speculate about possible implications of our findings.
As mentioned in the Introduction, the primary relaxation appears to be the driving force for the aging dynamics. Furthermore, it has been found by a thorough analysis of probe rotation and translation data that the width of the distribution of relaxation times appears to decrease for intermediate times during aging after a quench\cite{Thurau:2002p4151}. This finding is in accord with the fact that the distribution of activation energies shows the same time-dependence. 
Our results for an up-jump thus suggest that the width of the relaxation time distribution should broaden and then narrow again as a function of the time elapsed after the up-jump.

As another experimental technique that possbily can show a intermittant narrowing for a quench and a broadening for an up-jump is provided by measurements of the enthalpy relaxation following sudden temperature changes.
However, we have to point out here that at the same time the relaxation time decreases which may make it difficult to observe the effect of a broadening.

Other interesting implications of our results might arise when one compares different theoretical approaches to the glass transition problem that in equilibrium might give rise to very similar results.
We hope that our findings will stimulate calculations of the details of the aging dynamics for different models and probably help to discriminate among them.
\section*{Acknowledgment}
We thank Roland B\"ohmer and Gerald Hinze for fruitful discussions.
Financial support of the work in M\"unster by the Deutsche Forschungsgemeinschaft via the SFB 458 is acknowledged.\\
The centre for viscous liquid dynamics "Glass and Time" is sponsored by the Danish National Research Foundation (DNRF).

\begin{appendix}
\section*{Appendix: The trap model}
\setcounter{equation}{0}
\renewcommand{\theequation}{A.\arabic{equation}}
The trap model was introduced into glass science in 1987\cite{DYRE:1987p2799}.
The temporal evolution of the population of a considered trap is determined by the escape rate $\k_T(\e)=\k_\infty e^{\b\e}$ which depends only on the trap-energy.
(The rate $\k_\infty$ gives the overall time scale.
Throughout the calculations the time is measured in units of $\k_\infty$.)
After the escape from a given trap the destination trap is chosen at random.
In a continuous formulation, the populations obey the
following master equation:
\be\label{ME.p}
{\dot p}_T(\e,t)= -\k_T(\e)p_T(\e,t)+\rho(\e)\int\!d\e'\k_T(\e')p_T(\e',t) 
\ee
We have used the subscript $T$ to emphasize that $p_T(\e,t)$ depends on temperature 
$T=1/\b$.
Throughout the calculations the DOS is chosen to be Gaussian
\be\label{DOS.Gauss}
\rho(\e)\!=\!{1\over\sqrt{2\pi}\s}e^{-\e^2/(2\s^2)}
\ee
with $\s=1$. 
This choice for the DOS guarantees that the system reaches equilibrium at all temperatures
(temperatures are measured in units of $\s$). 
The corresponding equilibrium populations at a given temperature $T$ are found to be Gaussian: 
\be\label{peq.e}
p^{\rm eq}_T(\e)={1\over\sqrt{2\pi}\s}e^{-(\e-{\bar\e}_T)^2/(2\s^2)}
\quad\mbox{with}\quad {\bar\e}_T=-\b \s^2
\ee
The solution of eq.(\ref{ME.p}) allows the computions of all quantities of interest.
For instance, the population of the trap with trap energy $\e$ is given by
\be\label{pop.t}
p_T(\e,t)=\int\!d\e_0G_T(\e,t-t_0|\e_0)p(\e_0,t_0)
\ee
where $p(\e_0,t_0)$ denotes the population at the initial time $t_0$ and 
$G_T(\e,t-t_0|\e_0)$ is the conditional probability to find the system in trap $\e$ at time $t$ provided it started in $\e_0$ at $t_0$.
The initial population is usually taken as the equilibrium population at the initial temperature $T_0$, 
\be\label{p.0.eq}
p(\e_0,t_0)=p^{\rm eq}_{T_0}(\e_0),
\ee
cf. eq.(\ref{peq.e}).
From eq.(\ref{pop.t}), one can obtain all further quantities like the observables, for example the moments of the energy distribution
\be\label{E.n.t}
E^n_T(t)=\int\!d\e \e^np_T(\e,t).
\ee
We solely consider instantaneous changes of temperature $T_0\to T$, changes with a finite cooling or heating rate can be considered in a similar way, by assuming constant temperatures in small time-intervals. 
As the populations reach equilibrium for long times, 
$p(\e,t\to\infty)=p^{eq}_{T}(\e)$, we find after a temperature jump from $T_0$ to $T$ that the energy evolves according to eq.(\ref{E.n.t}) with $n=1$ and that 
$E(t=0)=E_{T_0}^{\rm eq}=-\b_0\s^2$ and the final value $E(\infty)=E_{T}^{\rm eq}=-\b\s^2$.
\end{appendix}
\end{document}